\documentclass
[aps,twocolumn,preprintnumbers,amsmath,amssymb,superscriptaddress]{revtex4}%
\usepackage{amsfonts}
\usepackage{amsmath}
\usepackage{amssymb}
\usepackage{graphicx}
\usepackage{textcomp}%
\providecommand{\U}[1]{\protect\rule{.1in}{.1in}}

\begin{document}
\title{Universal Dynamics of Kerr Frequency Comb Formation in Microresonators}
\author{T.~Herr}
\affiliation{\'{E}cole Polytechnique F\'{e}d\'{e}rale de Lausanne (EPFL), CH~1015,
Lausanne, Switzerland}
\author{J.~Riemensberger}
\affiliation{\'{E}cole Polytechnique F\'{e}d\'{e}rale de Lausanne (EPFL), CH~1015,
Lausanne, Switzerland}
\author{C.~Wang}
\affiliation{Max-Planck-Institut f\"{u}r Quantenoptik, 85748~Garching, Germany}
\author{K.~Hartinger}
\affiliation{Menlo Systems GmbH, 82152~Martinsried, Germany}
\author{E.~Gavartin}
\affiliation{\'{E}cole Polytechnique F\'{e}d\'{e}rale de Lausanne (EPFL), CH~1015,
Lausanne, Switzerland}
\author{R.~Holzwarth}
\affiliation{Menlo Systems GmbH, 82152~Martinsried, Germany}
\author{M.~L.~Gorodetsky}
\affiliation{Faculty of Physics, Moscow State University, Moscow 119991, Russia}
\author{T.~J.~Kippenberg}
\email{tobias.kippenberg@epfl.ch}
\affiliation{\'{E}cole Polytechnique F\'{e}d\'{e}rale de Lausanne (EPFL), CH~1015, Lausanne, Switzerland}
\affiliation{Max-Planck-Institut f\"{u}r Quantenoptik, 85748~Garching, Germany}

\begin{abstract}
Optical frequency combs allow for precise measurement of optical frequencies and are used in a growing number of applications beyond spectroscopy and optical frequency metrology. A class of compact microresonator based frequency comb generators has emerged recently based on (hyper)-parametric frequency conversion, mediated by the Kerr-non-linearity, of a continuous wave laser beam. Despite the rapid progress and the emergence of a wide variety of micro-resonator Kerr-comb platforms, an understanding of the dynamics of the Kerr comb formation is still lacking. In particular the question in which regime low phase noise performance can be achieved has so far not been answered but is of critical importance for future application of this technology. Here an universal, platform independent understanding of the Kerr-comb formation dynamics based on experimental observations in crystalline MgF$_{2}$ and planar Si$_3$N$_4$ comb generators is given. This explains a wide range of hereto not understood phenomena and reveals for the first time the underlying condition for low phase noise performance. 
\end{abstract}
\maketitle
 
\section{Introduction}
Optical frequency combs \citep{Holzwarth2000,Jones2000a,Udem2002,Cundiff2003} have revolutionized the field of frequency metrology and spectroscopy. By providing accurate frequency markers they allow for precise measurement of optical frequencies and are enabling ingredient to a variety of applications \citep{Newbury2011}. A class of monolithic frequency comb generators was discovered recently \citep{Del'Haye2007} by coupling a continuous wave (CW) laser to a high finesse fused silica microcavity \citep{Armani2003}. Here, the high intensities due to the tight spatial and long temporal confinement of the electromagnetic field, in combination with the material's Kerr-non-linearity allow for parametric frequency conversion\citep{Kippenberg2004a,Savchenkov2004}, particularly (cascaded) four-wave-mixing (FWM), resulting in an optical frequency comb\citep{Del'Haye2007}. These Kerr-combs could complement conventional frequency combs in applications where high power per comb line (typically several hundreds of micro-Watt) and high repetition rate (i.e. frequency spacing between the comb lines of typically 10-1000~GHz) are desirable, such as astronomical spectrometer calibration \citep{Murphy2007,Steinmetz2008,Li2008}, direct comb spectroscopy \citep{Diddams2007}, optical arbitrary waveform generation \citep{Jiang2007, Weiner2011}, and advanced telecommunications (For a recent review see ref.~\citep{Kippenberg2011}). Kerr-comb generation using optical microresonators in this manner has been demonstrated in a variety of other resonator host materials and geometries including crystalline CaF$_2$ \citep{Savchenkov2008a, Grudinin2009}, MgF$_2$ \citep{Herr2011,Liang2011,Wang2011} resonators, fused silica microspheres \cite{Agha2007}, planar high-index silica \citep{Razzari2010} and Si$_3$N$_4$ ring-resonators \citep{Levy2010,Foster2011}, and compact fibre cavities \citep{Braje2009}. Over the past years significant advancement of the Kerr-comb technology has been achieved by demonstrating a fully phase stabilized optical Kerr-comb \citep{Del'Haye2008}, generation of octave spanning spectra, both in fused silica microresonators \citep{Del'Haye2011} as well as planar Si$_3$N$_4$ micro-ring-resonators \citep{Okawachi2011}, the detection and shaping of a pulsed output spectrum \citep{Weiner2011}, and the extension of spectral coverage towards the visible \citep{Savchenkov2011} (albeit with very low spectral coverage of $<5$~nm) and the mid-infrared spectral region \citep{Wang2011}. Despite these advances however, in all prior work a broad spectrum (i.e. several hundreds of nanometers in span), with a low repetition rate (below $100$~GHz as required for direct detection and stabilization), and good phase noise performance (i.e. well defined, narrow comb lines) have not been achieved simultaneously, but are essential ingredients to bring this new technology to maturity \citep{Kippenberg2011}. 

In fact, while good noise performance was demonstrated in early work (e.g. ref.~\citep{Del'Haye2007}) recent work aiming at broader spectra and lower repetition rate reported linewidth broadening in octave spanning spectra \citep{Del'Haye2011} and multiple radio frequency (RF) beat-notes in low repetition rate \citep{Papp2011} fused-silica Kerr-comb systems. However, these phenomena are not described by current Kerr-comb theory (e.g. \citep{Chembo2010}). As the RF beat-note provides a means of detecting and stabilizing the comb's repetition rate, an understanding of the underlying processes responsible for multiple and broad RF beat-notes is an outstanding scientific challenge and essential to further advance the Kerr-comb technology. 
 
Here we unveil and describe for the first time the origin of broad and multiple RF beat-notes, and attribute this to an universal platform independent behaviour in Kerr-comb generators of vastly different material and geometry. Moreover, we provide a quantitative analysis of the regime of low phase noise operation. As such our results constitute an important step towards low phase noise comb generators.

\section{Experimental Systems}
\begin{figure*}[ptbh]
\begin{center}
\includegraphics[width=0.8\textwidth]{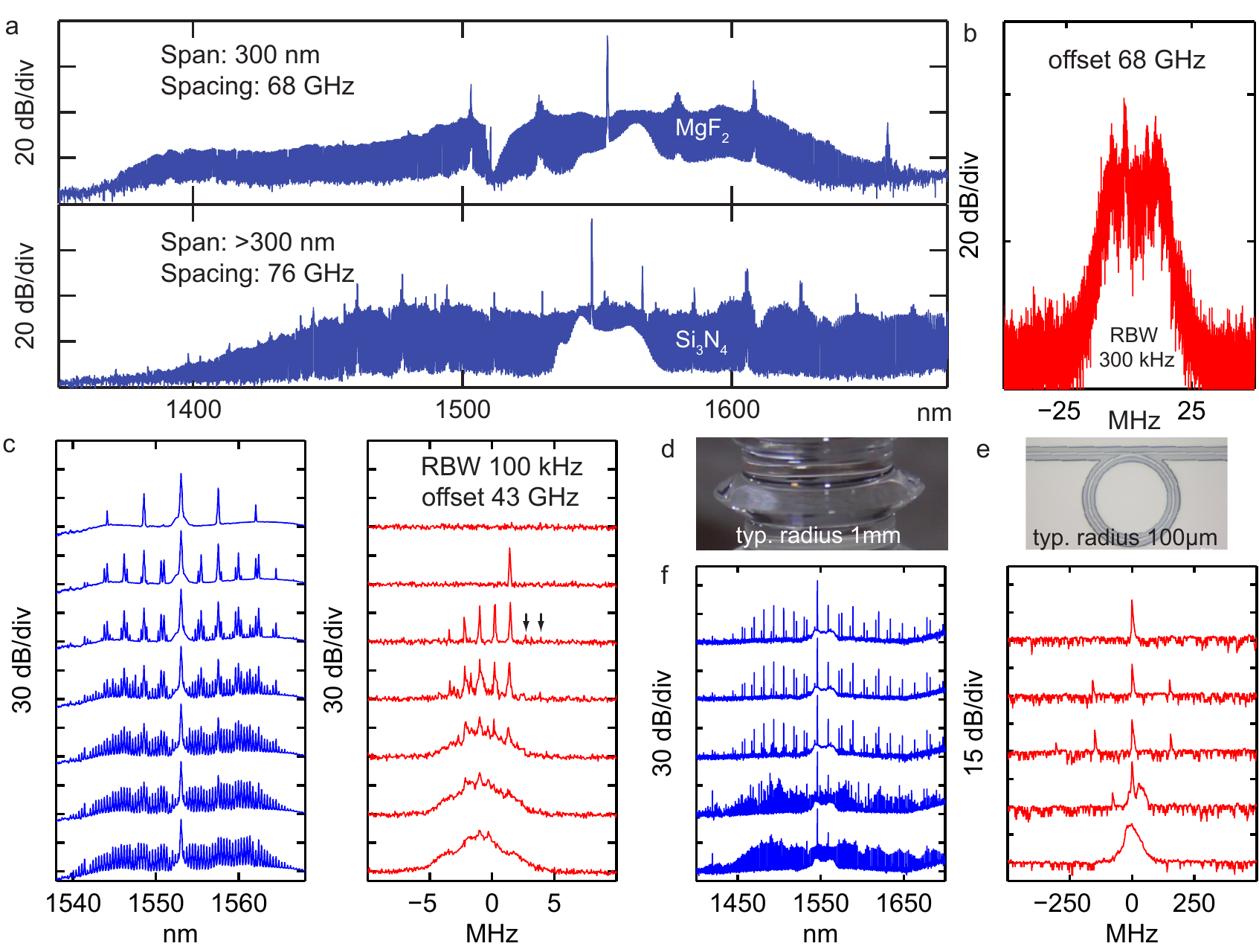}
\end{center}
\caption{\textbf{a.} Optical frequency comb generation in crystalline MgF$_{2}$ (500~mW pump power, 1~MHz Linewidth) and Si$_3$N$_4$ micro-resonators (3~W pump power, 200~MHz Linewidth) with a repetition rate of 68~GHz and 76~GHz, respectively. \textbf{b.}Both spectra exhibit broad radio-frequency (RF) beat-notes, shown here for the case of the upper spectrum. \textbf{c.} Evolution of the optical frequency comb spectrum (blue) generated in a 43~GHz MgF$_2$ resonator while tuning the pump laser into resonance and simultaneously recorded RF beat-note (red). \textbf{f.} same as c for a 200~GHz Si$_3$N$_4$ ring-resonator. The measurement of the RF beat-note is replaced by an equivalent beat-note measurement of a comb line and an independent external continuous wave laser. \textbf{d,e.} Optical images of the resonators used in c and f, respectively.}
\label{phenofig}%
\end{figure*}

We employ two entirely different systems, namely crystalline MgF$_{2}$ resonators \citep{Herr2011,Liang2011,Wang2011} (resonance width $\sim1$~MHz)  and Si$_3$N$_4$ micro-ring resonators \citep{Levy2010} (resonance width $\sim200$~MHz) for Kerr-comb generation with a CW pump laser evanescently coupled to the resonator (cf. Fig~\ref{phenofig}d,e). In case of the MgF$_{2}$ resonators a tapered fibre is used for coupling, whereas the coupling to the Si$_3$N$_4$ rings is achieved via an on-chip waveguide, which itself is interfaced by lensed fibres. The comb emerges when a blue-detuned pump laser is gradually tuned into resonance thereby increasing the circulating power inside the resonator. Parametric frequency conversion sets in once the parametric threshold \citep{Kippenberg2004a} is reached. In this state the resonator is thermally locked \citep{Carmon2004} to the blue-detuned pump laser. In Fig.~\ref{phenofig}a it is shown that Kerr-combs can generally satisfy the opposing criteria of low repetition ($\lesssim100$~GHz) rate and large spectral coverage, by the demonstration of more than $300$~nm wide spectra with a line spacing of $68$~GHz in a MgF$_{2}$ and $76$~GHz in a Si$_3$N$_4$ system. 
However, similar to  previous work in fused-silica microresonators \citep{Del'Haye2011} a broad RF beat-note is observed, in both, the MgF$_{2}$ and the Si$_3$N$_4$ systems (setup as shown in Fig.~\ref{exp1fig}g).With two experimental platforms at hand that despite their differences in material and geometry show qualitatively identical phenomena, platform independent conclusions can be drawn based on detailed analysis and comparison of the two systems.

\section{Multiple and Broad Beat Notes}
In principle, a multitude of physical processes can account or contribute to the observation of broad RF beat-notes. These processes include line broadening phase noise mechanisms, such as thermorefractive noise, thermoelastic noise, thermal Brownian motion, ponderomotive noise, photothermal noise, laser phase noise and self/cross phase modulation \cite{Braginsky1999, Gorodetsky2004, Matsko2007, Del'Haye2011}. Phase noise can also be generated by optomechanical coupling \citep{Rokhsari2005, Ma2007, Savchenkov2011a} of optical and mechanical resonator modes, by thermal oscillations of the resonator \citep{Fomin2005}, as well as by the interaction of light with higher order transverse modes \citep{Savchenkov2011}. 

The high number of physically entirely different processes makes it difficult to identify the mechanisms responsible for the broad RF beat-notes. To shed light into the origin of the brfoad RF beat-note, we investigate the beat-note at different stages during the comb formation, i.e. for different values of the laser detuning (For practical reasons the RF beat note measurement is replaced by recording the beat-note of a CW laser with a strong comb line in case of the Si$_{3}$N$_{4}$ system). Fig.~\ref{phenofig}c,f reveal an intriguing behaviour in both SiN and crystalline systems. The broad beat-note, visible in the lowermost data set, in fact consists of a discrete spectrum of multiple narrow beat-notes, whose number is increasing as more power is coupled to the resonator (cf. Fig.~\ref{phenofig}c,f). This result \textit{unifies} the observation of broad and multiple beat-notes made individually in earlier work \citep{Del'Haye2011,Papp2011}. Importantly however, this phenomenology prevails, both, in the crystalline MgF$_{2}$ whispering gallery mode resonator, as well as in the Si$_{3}$N$_{4}$ system despite their vastly different geometries and different material characteristics. The striking similarity between the two comb generators shows the universality of the phenomenon, which is not restricted to a particular platform. By interpreting broad and multiple beat-notes as identical phenomena, pure line broadening phase noise mechanisms (as those mentioned above) can be excluded as the principal reason for the observations. 

\begin{figure*}[ptbh]
\begin{center}
\includegraphics[width=0.8\textwidth]{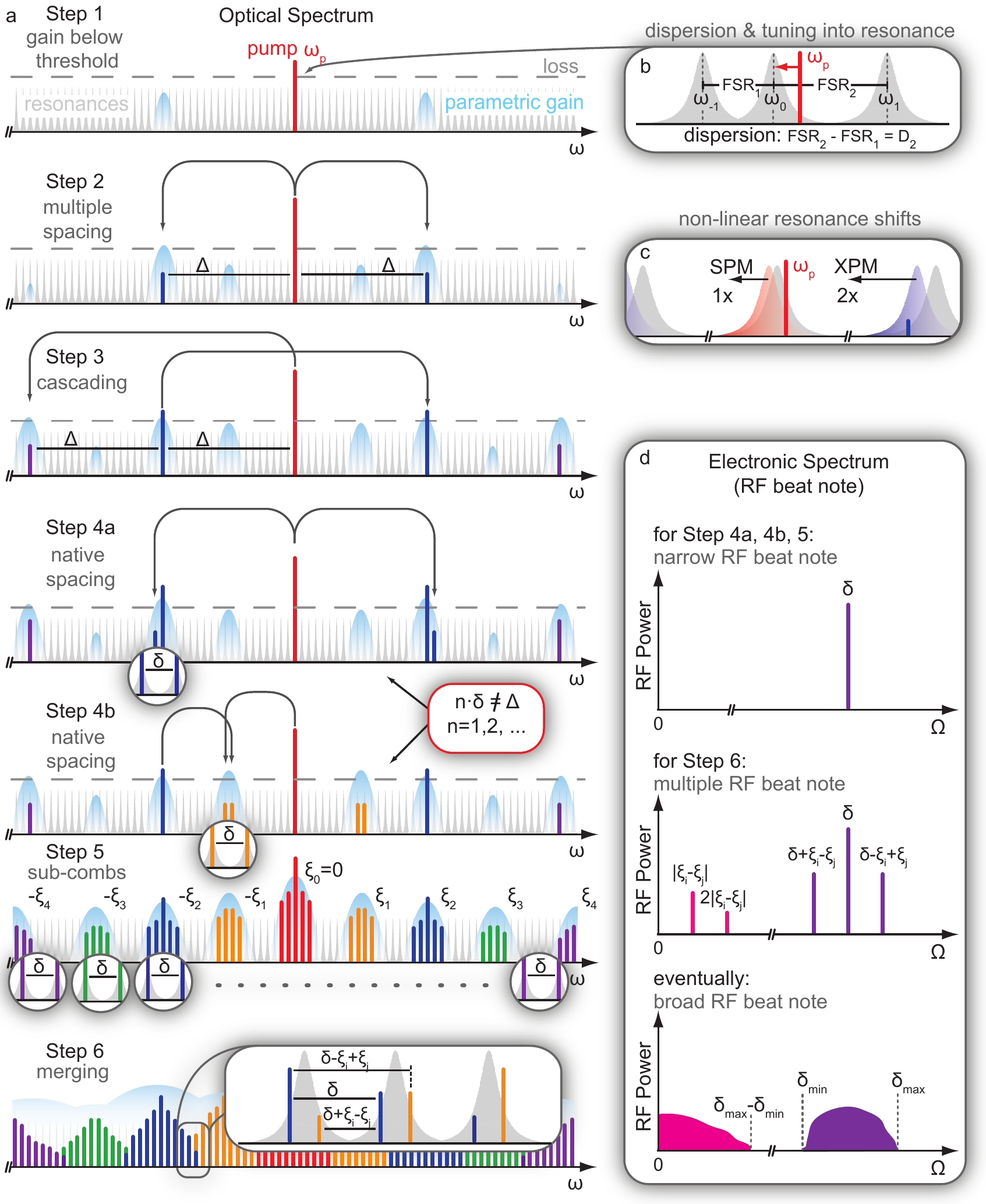}
\end{center}
\caption{\textbf{a. }Multiple mode spaced (MMS) Kerr-comb formation - \textbf{Step 1:} The parametric gain (indicated by the blue shaded gain lobes) is too small to overcome the cavity loss rate, i.e no lines are generated. \textbf{Step 2:} Initial degenerate four-wave mixing step in Kerr-comb generation. Generally, the frequency difference $\Delta$ between these primary lines corresponds to multiple mode numbers away from the pump mode. \textbf{Step 3:} Cascaded FWM generates more primary lines with the same spacing $\Delta$ introduced in Step 2. \textbf{Step 4a,b:} Secondary lines are generated around and/or in between the primary lines with a native spacing $\delta$, corresponding to neighbouring modes. The spacings $\Delta$ and $\delta$, however, are not commensurate i.e. $\Delta$ is not an integer multiple of $\delta$. \textbf{Step 5:} Spectrally separated sub-combs form around the primary lines and the lines generated in Step 4b via non-degenerate FWM. All sub-combs i have the same spacing $\delta$, but different offset frequencies $\xi_\mathrm{i}$. \textbf{Step 6:} The sub-combs merge and form a gap-free spectrum of lines, where more than one line can exist in a single resonator mode. \textbf{b. } Dispersion of the resonator and blue detuned pumping. \textbf{c. } Self-(SPM) and Cross-Phase (XPM) Modulation induced by a strong pump on the pumped resonance (red) and other cavity resoances (blue), respectively. \textbf{d. } RF signal observed when sending the generated Kerr-comb spectrum on a photo-detector, exhibiting single and multiple narrow RF peaks, as well as broadband signals.}%
\label{schemeFig}%
\end{figure*}

Moreover, while recently frequency combs on transversal order modes in crystalline resonators have been reported \citep{Savchenkov2011}, these multi-mode effects are unlikely given the results of broadband resonator spectroscopy following ref.~\citep{Del'Haye2009}, showing that, except for mode crossings, the resonances are generally separated by several GHz. This is particularly true for the Si$_3$N$_4$ resonators studied in this work. Interaction via FWM between different mode families (which generally posses different FSRs) is suppressed by the requirement of angular momentum conservation. 
While crystalline resonators support mechanical radial breathing modes \citep{Hofer2010}, as well as surface acoustic modes \citep{Savchenkov2011a,Bahl2011}, which could equally give rise to multiple, modulation beat-notes and chaotic opto-mechanical oscillations \citep{Carmon2007} in crystalline resonators the planar, on-chip geometry of the Si$_{3}$N$_{4}$ system (which has no free boundaries due to the oxide embedded resonators) render thermal and mechanical oscillations implausible. Instead we consider here a novel line of reasoning that has so far not been considered and suggest, that the cause for the broad and multiple beat-notes lies in the \textit{dynamics} of Kerr-comb formation. Generally, we observe a pronounced hysteresis (also see ref.~\citep{Okawachi2011}) when tuning the laser into and out of resonance in the sense that more than one stable comb state can be observed for a specific pump laser detuning. This emphasizes the role of the dynamics of the Kerr-comb formation and shows that the temporal order in which comb lines form matters.

\begin{figure*}[ptbh]\begin{center}
\includegraphics[width=0.8\textwidth]{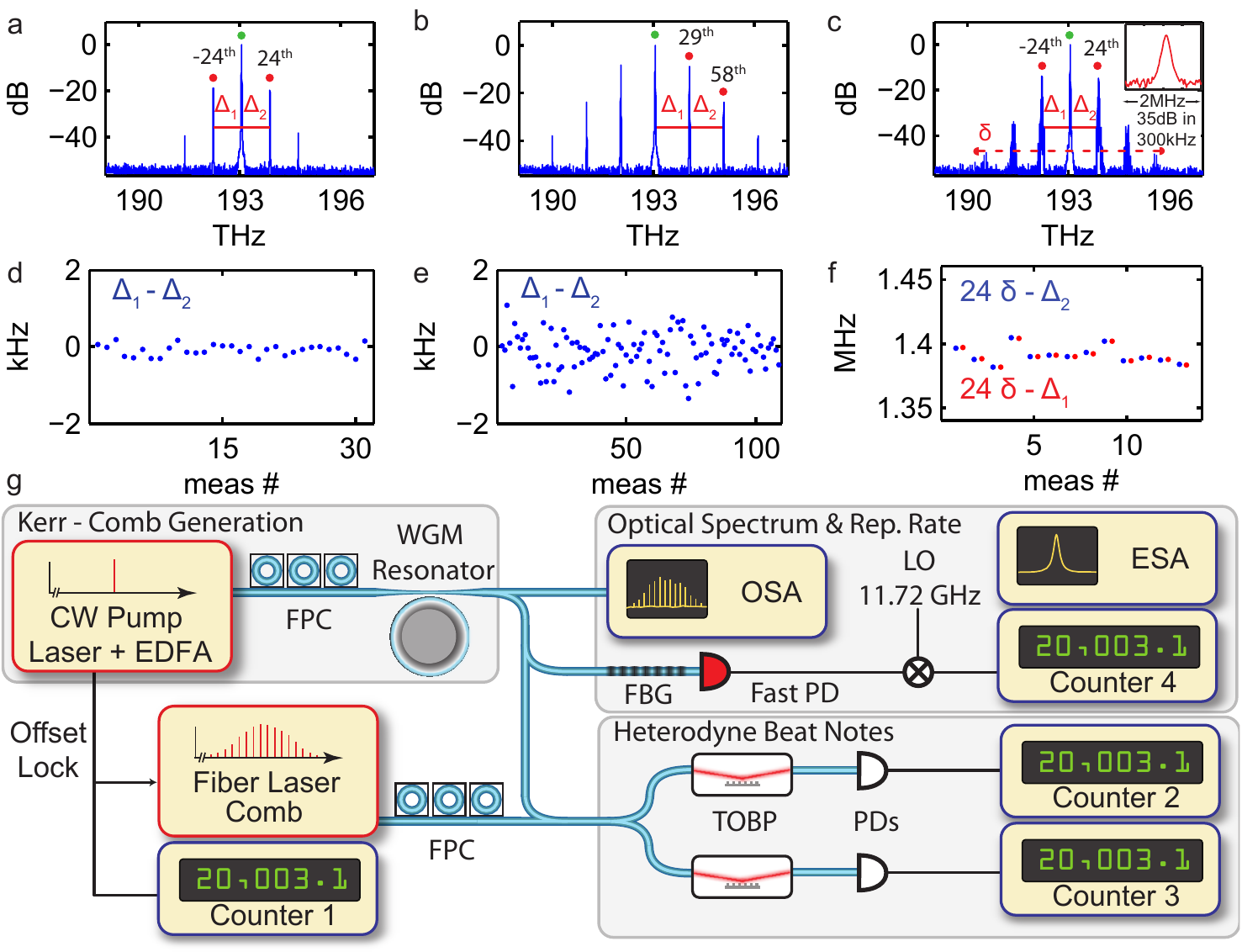}
\end{center}
\caption{\textbf{a, b, c.} Optical frequency comb spectra generated in a 35~GHz MgF$_2$ resonator using 80~mW of pump power. In a,b only primary comb lines with a spacing corresponding to 24 and 29 free spectral ranges (FSR) are generated. In c additional comb lines spaced by a single FSR are generated around the primary spaced lines allowing for direct detection of the native spacing $\delta$ using a fast photo-diode. The resulting RF signal consists of a single and well defined line (cf. Inset) with a signal-to-noise of $35$~dB within a resolution bandwidth of $300$~kHz. Multi-heterodyne beat-notes between select lines in the generated spectra and a conventional 250~MHz fibre laser comb are recorded and sent to frequency counters. The respective lines are marked with dots (green=pump, red= generated comb line). Frequency counting of the beat-notes allows for precise determination of the frequency distances $\Delta_\mathrm{1}$ and $\Delta_\mathrm{2}$. \textbf{d.} Difference of $\Delta_\mathrm{1}$ and $\Delta_\mathrm{2}$ as indicated in a. No deviation from the symmetry of the primary comb lines  with respect to the pump is detected. \textbf{e.} Difference of $\Delta_1$ and $\Delta_2$ as indicated in b. No deviation from the equidistance of the primary comb lines is detected. \textbf{f.} The spacing $\Delta_\mathrm{1}$ and $\Delta_\mathrm{2}$ are not commensurate with the native spacing $\delta$ between neighbouring comb lines, i.e. $\Delta_\mathrm{1}$ and $\Delta_\mathrm{2}$ are not integer multiples of $\delta$, but show a mismatch of approximately 1.4~MHz. \textbf{g.} Setup used for frequency comb generation, detection of the optical spectrum and RF beat-note spectrum, as well as frequency counting of the RF beat-note, and multi-heterodyne beat-notes of the pump laser and two select Kerr-comb lines with a 250~MHz fibre laser comb. EDFA = Erbium Doped Fiber Amplifier, FPC = Fiber Polarization Controller, OSA = Optical Spectrum Analyzer, ESA = Electronical Spectrum Analyzer, FBG = Fiber Bragg Grating, LO = Local Oscillator, PD = Photodiode, TOBP = Tunable Optical Bandpass Filter.}
\label{exp1fig}
\end{figure*}

\section{Kerr-Comb Formation}
Fig.~\ref{schemeFig}a explains schematically the initial steps of Kerr-comb formation. Generally the first comb lines are generated in a degenerate FWM process symmetrically to the pump frequency, as soon as the parametric gain overcomes the loss of the cavity (cf. fig.~\ref{schemeFig}a, Step1,2). The distance in terms of relative mode number $\mu$ of the new lines to the pump depends on the pump power and the dispersion of the resonator, which affects the cold resonance frequencies:  
\begin{equation}
\omega_{\mathrm{\mu}} = \omega_{\mathrm{0}} + D_{\mathrm{1}} \cdot \mu + \frac{1}{2}D_{\mathrm{2}} \cdot \mu^2 + ... 
\end{equation}
where $\omega_\mathrm{0}$ is the frequency of the pump mode $\mu=0$, and  $D_{\mathrm{1}}$ corresponds to FSR of the resonator and $D_{\mathrm{2}}$ to the difference of the two FSRs adjacent to the expansion frequency $\omega_{\mathrm{0}}$ (cf. fig~\ref{schemeFig}b). Note that $D_\mathrm{2}$ relates to the group velocity dispersion (GVD) via $\beta_2 = -2\pi D_\mathrm{2}/D_\mathrm{1}^3$. It has been observed, that the first lines oscillate either in the resonator mode directly adjacent to the pump (i.e. separated by approximately the FSR of the resonator) or a multiple thereof \citep{Savchenkov2008,Agha2009, Del'Haye2009b,Razzari2010,Weiner2011}. Theoretically this can be understood within the framework of non-linear coupled mode equations as detailed later on and in the Supplementary Information (SI). In a simplified picture this can be explained by the non-linear mode-shifts induced by the pump laser via self- and cross-phase modulation compensating the dispersion $D_2$ (cf. fig.~\ref{schemeFig}c).

Based on these experimental observations, we distinguish two scenarios of comb formation. We refer to these scenarios as \textit{multiple} mode spaced (MMS) and \textit{natively} mode spaced (NMS) combs. 
In MMS combs the first parametric side-bands are generated at a spacing $\Delta$, corresponding to a \textit{multiple} mode number difference, away from the pump (cf. Fig~\ref{schemeFig}a, Step 2 and experimentally fig.~\ref{phenofig}c,f and \ref{exp1fig}a,b,c). Cascaded FWM transfers the initial spacing $\Delta$ to higher order side-bands (Fig.~\ref{schemeFig}a, Step 3). The initial spacing $\Delta$ is preserved and reproduced between all emerging lines, due to energy conservation in parametric processes \citep{Del'Haye2007}. We refer to these initial side-bands spaced by $\Delta$ as \textit{primary} comb lines. Importantly however, at a later stage of comb evolution, i.e. by increasing the coupled power natively $\delta$-spaced \textit{secondary} lines are observed to form around (Fig.~\ref{schemeFig}a Step 4a) and/or in between (Fig.~\ref{schemeFig}a Step 4b) the primary lines depending on e.g. coupling and pump power. Here, $\delta$ denotes the native spacing of the comb lines, corresponding to a mode number difference of one. The formation of lines with spacing $\delta$ can be understood in terms of the parametric gain lobes that are broad enough to allow two neighbouring modes to be populated (similar to multi-mode lasing). The position and width of the parametric gain lobes is illustrated in fig.~\ref{schemeFig} and theoretically calculated in the SI. The spectrally separated sub-combs initiated in the processes shown in fig.~\ref{schemeFig}a, Step 4a,b grow via non-degenerate FWM when the power coupled to the cavity is increased (cf. fig.~\ref{schemeFig}a, Step 5), and eventually merge to form a gap-free spectrum of lines. MMS combs have so far been observed in systems possessing a smaller FSR ($10-100$~GHz) or high pump power ($\gtrsim 1$~W) \citep{Savchenkov2008,Del'Haye2009b,Razzari2010,Weiner2011,Del'Haye2011}.  

In NMS combs the first parametric side-bands are generated at the native spacing $\delta$ of the resonator, i.e. in the modes $\mu = +1, -1$ adjacent to the pump \citep{Kippenberg2004a}. NMS combs can be interpreted as spacial case of MMS combs where $\Delta=\delta$. Consecutively, non-degenerate cascaded FWM leads to the generation of a frequency comb spectrum growing outwards from the pump, where the processes in fig.~\ref{schemeFig}a, Step 4a,b are not possible. 

Note that the described formation process of the comb is not only observed experimentally but is also in agreement with numerical modelling based on non-linear coupled mode equations (in preparation \citep{ArchiveGHK}).

In the following we will investigate the properties of MMS combs to understand if and how their formation mechanism affects the width and number of the RF beat-notes. To do so, we perform a multi-heterodyne experiment between a Kerr-comb generated in a MgF$_{2}$ resonator with a FSR of $35$~GHz and a conventional stabilized mode locked fibre laser comb (Menlo Systems GmbH) with a repetition rate of $250$~MHz. As detailed in the Methods section \ref{methods:multhet} this allows to accurately determine the position of select Kerr-comb lines (Fig.~\ref{exp1fig}). In first measurements the symmetry of the two innermost MMS side-bands with respect to the pump and the equidistance between adjacent MMS side-bands(Fig.~\ref{exp1fig}a,b,c,d) is verified to the sub-kHz level. 

More importantly, however, we address in a second experiment (Fig.~\ref{exp1fig}c,f) the vital question of the commensurability of \textit{primary} and \textit{secondary} comb lines, that form when the laser is tuned further into resonance. More precisely, it is tested whether the primary spacing $\Delta$ is an integer multiple of the native spacing $\delta$, i.e. whether the generated optical spectrum forms an intermittent but consistent frequency comb, whose gaps can be filled by increasing the coupled power and extending the existing sub-combs. Strikingly, we find that this is \textit{not} the case. A mismatch of $\mu\delta-\Delta_{\mathrm{1,2}}=1.4 \mathrm{~MHz},\ (\mu=24)$ between the actual position of the primary comb line and the expected position assuming a comb line grid with spacing $\delta$ centred on the pump laser frequency is observed. This mismatch exceeds the precision of the measurement by far. The redundancy in performing this measurement with a pair of symmetric primary comb lines rules out artefacts in the frequency counting such as cycle slips that would lead to an incorrectly detected mismatch. Importantly, the RF beat-note resulting from the mutual beating between neighbouring natively $\delta$-spaced comb lines is found well defined and unique over the full spectral width of the comb spectrum (cf. Inset in Fig.~\ref{exp1fig}c). Based on these observations, we conclude that the optical spectrum consists of multiple equidistant combs $\mathrm{i}$, composed of  lines $\mu$, at optical frequencies $\omega_{\mu}^{\mathrm{(i)}}$ separated by the the same spacing $\delta$:
\begin{equation}
\omega_{j}^{\mathrm{(i)}}=\xi_{\mathrm{i}}+j\cdot \delta,\ j=0,\pm1,\pm2,...
\end{equation}
These equidistant combs, however, posses different, non commensurate offset frequencies $\xi_{\mathrm{i}}\neq\xi_{\mathrm{j}}+n\cdot\delta,\ i\neq j$ as illustrated in fig.~\ref{schemeFig}a, Step 5 (due to symmetry and equidistance of the primary lines $\xi_{\mathrm{0}}=0$ and $\xi_{\mathrm{-i}}=-\xi_{\mathrm{i}}$). 
Whenever isolated lines (i.e. not in the direct vicinity of existing lines) are generated (cf. fig.~\ref{schemeFig}a, Step 4b), new sub-combs with new offset $\xi_{\mathrm{x}}$ are defined, which then can grow as already described above. It is worth emphasizing, that even though all sub-combs are spectrally separated and reside at spectral positions with different resonator FSR, we found that their individual spacings $\delta$ are equal, resulting in a narrow and well defined RF beat-note (cf. fig.~\ref{schemeFig}d). This observation becomes clear when switching from a frequency to a time domain picture. Here the FWM can be interpreted as a time dependent modulation of the effective refractive index of the resonator. Assuming similar mode-profiles over the spectral width of the comb, this temporal refractive index modulation gives rise to phase modulation side-bands on on all modes, thereby, once generated, the spacing $\delta$ is transferred to \textit{all} spectral locations of the comb.

A critical point is reached, when the bandwidth growth of the sub-combs closes their spectral separation and leads to an overlap (Fig.~\ref{schemeFig}a, Step 6). As the combs start to merge, individual resonator modes are populated by multiple lines with slightly different optical frequencies. When monitoring the RF beat-note of $\delta$ on an electronic spectrum analyser (ESA),  symmetric side-bands around $\delta$ with frequencies $\delta_{\pm}^{\mathrm{i}}=\delta\pm\Delta\xi_{\mathrm{i}}$ with $\Delta \xi_{\mathrm{i}}=\xi_{\mathrm{i+1}}-\xi_{\mathrm{i}}$ being the offset difference between the two merging combs appear. This can be seen schematically in fig.~\ref{schemeFig}d or experimentally in Fig.~1c (third data set from top). Once these \textit{new} RF beat-notes are generated, they can spread throughout the optical comb spectrum, effectively leading to multiple lines within individual resonator modes and correspondingly even more RF beat-notes $\delta_{\pm}^{\mathrm{i,j}}=\delta\pm\Delta\xi_{\mathrm{i,j}}$ with $\Delta \xi_{\mathrm{i,j}}=\xi_{\mathrm{i}}-\xi_{\mathrm{j}}$. Eventually this leads to a broad RF beat note as indicated in fig.~\ref{schemeFig}d. Note that not all possible side beat-notes $\delta_{\pm}^{\mathrm{i,j}}=\delta\pm\Delta\xi_{\mathrm{i,j}}$ may be generated as the finite cavity bandwidth will suppress lines far off the resonance frequencies, thereby limiting the number of comb lines that populate a single resonator mode. Also note, that the shape of the broad RF beat-note emerging from multiple lines observed in Fig.~\ref{phenofig}c follows the (non-linearly broadened) line shape of the cavity resonance. The described pathway to multiple and broad RF beat-notes is absent in NMS combs, which is in agreement with previous experimental observation \citep{Del'Haye2007, Wang2011}.

\begin{figure*}[ptbh]
\begin{center}
\includegraphics[width=0.8\textwidth]{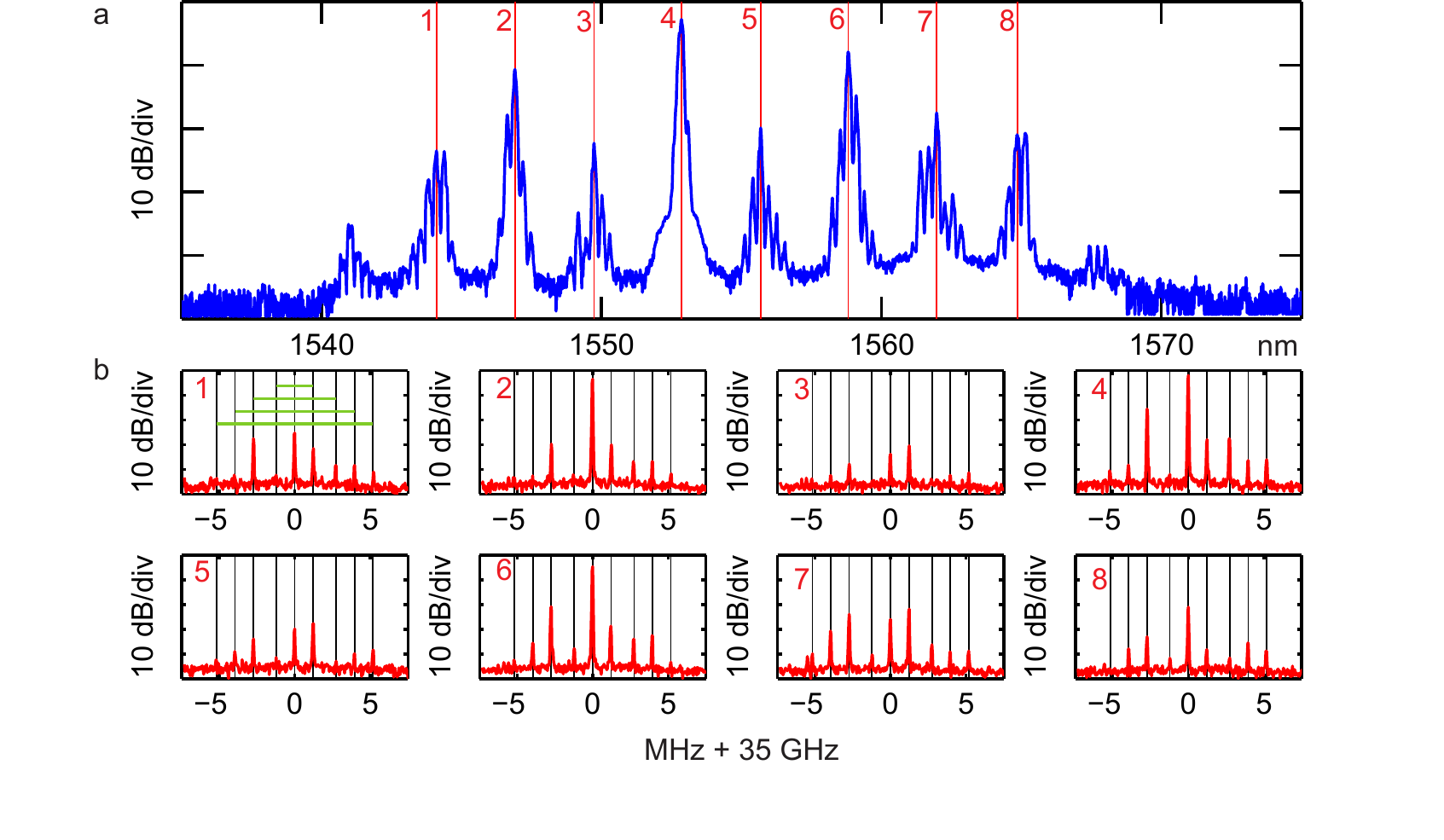}
\end{center}
\caption{\textbf{a.} Optical frequency comb spectrum generated in $35$~GHz MgF$_{2}$ resonator. The vertical lines mark eight different position where a narrow tunable bandpass filter was placed to filter out $0.8$~nm of the comb spectrum (corresponding to three comb lines). \textbf{b.} Radio frequency (RF) beat-note measured at the filtered portion of the spectrum using a fast photo-detector and resolution bandwidth of $300$~kHz.}
\label{repFiltered}%
\end{figure*}

To support this hypothesis experimentally, we measure the RF beat-note of a Kerr-comb not by sending the whole comb spectrum to the fast photo-detector, but by bandpass-filtering a narrow optical fraction ($0.8$~nm$\approx3$~FSR) prior to detection (cf. Fig.~\ref{repFiltered}). For the same comb, the bandpass filter is moved to eight different spectral positions and the RF beat-note between the filtered lines is measured. Several interesting novel observations can be made and explained by our hypothesis. First it can be seen that the number of detected RF beat-notes is higher than the number of filtered comb lines, evidencing that indeed more than one comb line exist in each cavity resonance. Note that this behaviour is surprising and contrary to the present understanding of Kerr combs  \citep{Chembo2010}. Second, by comparing the measured RF beat-note spectra at different spectral position of the filter, we are able to demonstrate that the same RF beat-note frequencies are present (with varying amplitude) at essentially \textit{all} spectral regions of the comb, as discussed above. Generally, as shown in the SI (cf. Fig.~\ref{combAndRep}) there is a correspondence between optical comb and RF beat-note spectrum. 

Importantly, our observations can also explain a further, hereto unexplained phenomenon observed in MMS combs (in a state corresponding to Step 6 in fig.~\ref{schemeFig}a). If the transmission of the pump laser is recorded, oscillatory features become apparent, whose emergence coincide with the generation of the frequency comb. Indeed, while the spacings $\delta$, $\delta_{\pm}^{\mathrm{i}},$, and $\delta_{\pm}^{\mathrm{i,j}}$ are close to the FSR of the resonator and therefore require dedicated RF equipment for their detection, this phenomenon finds its equivalent in the low RF regime, where the frequencies $\Delta \xi_{\mathrm{i}}$ and $\Delta\xi_{\mathrm{i,j}}$ can be measured in the transmission spectrum (cf. Fig~\ref{schemeFig}d and experimentally in the SI, Fig.~\ref{repAndTransLowRF}). In particular, broad RF beat-notes at frequencies corresponding to the native spacing will coincide with a broad signal at low frequencies close to DC. This provides an easy access to noise analysis of Kerr-combs, without the need for dedicated RF equipment. In the light of the understanding presented here, the observation made recently in a Si$_ 3$N$_4$ ring-resonator  \citep{Okawachi2011} of a so far unexplained broadband rise of the noise floor (measured up to $25$~MHz), as well as a noise peak at $11$~MHz (which was attributed to relaxation oscillations of the pump laser) may also be explained as the low frequency signal corresponding to broad and/or multiple RF beat notes between natively spaced lines, similar to the ones observed here (cf. Fig.~\ref{phenofig} and SI, Fig.~\ref{repAndTransLowRF}).

\begin{figure*}[ptbh]
\begin{center}
\includegraphics[width=0.8\textwidth]{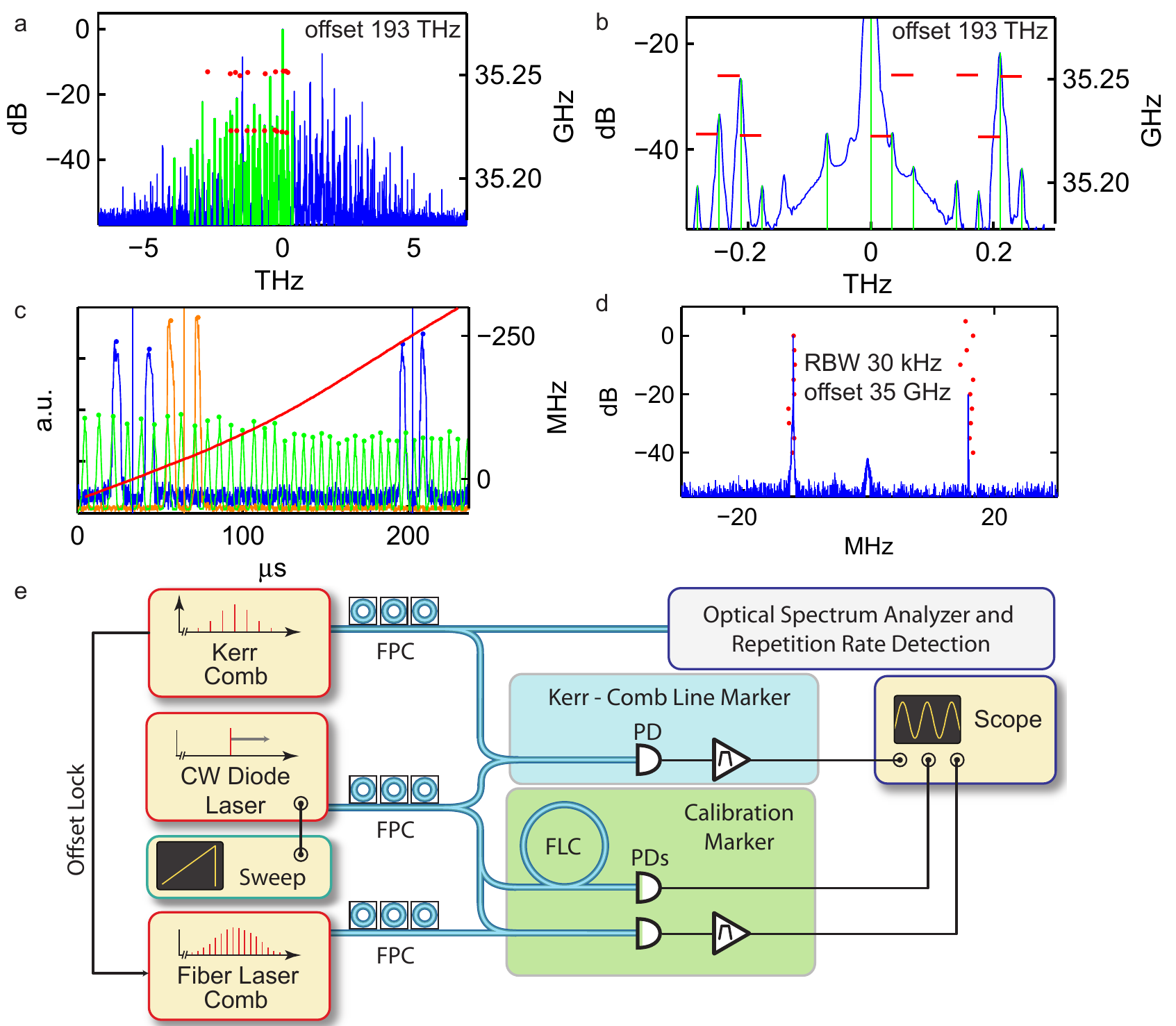}
\end{center}
\caption{\textbf{a.} Optical frequency comb spectrum (blue) and reconstructed comb lines (green). The difference frequency between neighbouring comb lines is shown via red markers, drawn between the respective comb lines. The height of the marker, corresponds to the frequency difference between the neighbouring lines, which can be read from the right vertical axis. \textbf{b.} Zoom into (a) in the pump laser region. \textbf{c} Raw data showing the beat-note markers between diode laser and fibre laser comb (blue, normalized absolute values), between diode laser and Kerr-comb (orange, normalized absolute value), and the resonances of the diode laser scanning over the fibre loop cavity (green, normalized absolute value of inverted signal). The lines of the fibre laser comb and the Kerr-comb are reconstructed in the center of the blue and orange markers. The detuning (red, right vertical axis) of the scanning diode laser can be inferred relative to a fibre laser comb line by means of the fibre loop cavity resonances. Based on this detuning curve a relative detuning is assigned to the Kerr-comb line, which then allow to measure the frequency difference to other Kerr-comb lines. \textbf{d} The direct measurement of the RF beatnote (blue), as the beat-note of neighbouring comb lines and the frequency difference between neighbouring reconstructed comb lines (red dots, arbitrary vertical position) does agree well. \textbf{e. } Setup for high-resolution broadband Kerr-comb reconstruction. A continuous wave diode laser scanning over a span of approximately $4$~THz is separated into three beams. One beam is sent to a fibre loop cavity and the transmission is detected on a photo-diode, the other two beams are individually combined with a fibre laser comb and a Kerr-comb spectrum respectively. The beat-notes between the diode laser beams and the respective combs are detected by photo-diodes. The resulting electronic signal is filtered using $10$~MHz bandpass filters (bandwidth approximately 2~MHz). While the diode laser is scanning, the signal of all photo-diodes is recorded simultaneously on an oscilloscope and saved for later analysis. For convenience the Kerr-comb pump laser is offset locked to the fibre comb, which allows for a straightforward identification of the reconstructed pump line by its relative detuning.}
\label{exp2fig}%
\end{figure*}

In order to prove, that the interpretation of the RF beat-note as the difference frequency between the generated comb lines holds true (as opposed to e.g. side-bands generated by mechanical or thermal modulation) we perform a high resolution spectroscopic experiment of the generated comb (Fig.~\ref{exp2fig}) with sub-MHz-resolution over a span of $4$~THz. This novel method allows a line-by-line reconstruction of the Kerr-comb, i.e. measuring the exact frequency of each comb line. The experimental setup is shown in Fig~\ref{exp2fig}e and explained in detail in the Methods section \ref{methods:combreco}. Following the approach in ref.~\citep{Del'Haye2009} a diode laser is scanned over the spectral region of interest, while using the regular frequency markers provided by the band-pass filtered RF signal of its beat-note with a fibre laser comb for calibration. In the same manner the beat-notes between the scanning laser and Kerr-comb are used to determine the positions of the Kerr-comb lines. To correct for changes in the scan speed of the laser additional calibration markers are provided by also recording the transmission signal of the laser passing by fibre-loop cavity with a FSR much smaller than the repetition rate of the fibre laser comb. We estimate the precision of the method to $\leq 10\%$ of the fibre-loop FSR, i.e below $1$~MHz. After successful reconstruction (cf.Fig.~\ref{exp2fig}a,b) of the relative Kerr-comb line frequencies, the differences frequency between these lines is calculated. 
The spacings between the reconstructed comb line frequencies agrees perfectly with the RF beat-note signal from the direct measurement (Fig.~\ref{exp2fig}d). Importantly, this experiment not only allows to find the RF beat-note frequencies but also the position within the optical spectrum of the comb where they are generated. It can be seen that the two dominating RF beat-notes indeed correspond to the differences between neighbouring Kerr-comb lines and that they are evenly distributed throughout the comb spectrum (cf. Fig.~\ref{exp2fig}a,b). This reconstruction therefore provides further proof for the fact that the multiple beat-notes are generated by the Kerr-comb dynamics itself. 

\section{The Role of Dispersion}
In the following we show, that the dispersion parameter $D_\mathrm{2}$ is closely linked to the question, whether a comb evolves along the NMS or MMS pathway, and thereby impacts the RF beat-note. 

To estimate under which conditions NMS combs can be achieved, we determine the distance in terms of mode number between pump and first side-band (as done in previous work, e.g. refs.~\citep{Kippenberg2004a, Matsko2005, Agha2009, Chembo2010}) by solving the non-linear coupled mode equations and deriving the parametric gain for the initial degenerate FWM process, where we include the detuning of the pump laser, as well as the detuning of the generated side-bands with respect to the cavity resonances. The first side-bands are generated when the parametric threshold is reached, i.e. when the gain overcomes the cavity decay rate $\kappa$. For a constant launched pump power $P_{\mathrm{in}}$ and tuning the pump laser into resonance from the higher frequency side, the circulating power inside the cavity increases. At some point the parametric threshold is reached and the first four-wave mixing side-bands that are generated in the $\pm \mu_\mathrm{th}$ cavity resonance satisfy (for detail of the derivation see SI):	
 
\begin{equation}
\mu_{\mathrm{th}}\simeq\sqrt{\frac{\kappa}{D_\mathrm{2}}\left(\sqrt{\frac{8 \eta P_{\mathrm{in}} c n_\mathrm{2} \omega_\mathrm{p}}{\kappa^2 n_0^2 V_{\mathrm{eff}}}-1}+1\right)}
\nonumber\\
\label{qAtThreshold}
\end{equation}
Here, $\eta$ denotes the coupling strength($\eta=1/2$ for critical coupling), $\omega_p$ the pump frequency, $n_{\mathrm{0}}$ and $n_{\mathrm{2}}$ the refractive and non-linear indices, $V_{\mathrm{eff}}$ the effective non-linear mode volume, and finally, $c$ the speed of light. 

For a $35$~GHz MgF$_{2}$ system we obtain $\mu_\mathrm{th}\approx25$ ($\eta=1/2$, $\omega_{p}=193$~THz, $n_2=0.9 \times 10^{-20}$, $n_0=1.37$, $A_\mathrm{eff}=100 \times 10^{-12}$m$^2$, $\kappa=2\pi \times 10^6$~s$^{-1}$, $P_\mathrm{in}=0.1$~W, $D_2=10$~kHz), which agrees well with the experimental observation shown in fig.~\ref{exp1fig}a,b,c ($\mu_{\mathrm{th}}=24,29$). Similarly, this theory correctly predicts the $\mu_{\mathrm{th}}$ of the MMS combs observed in Si$_3$N$_4$ ring-resonators. The smallest possible $\mu_\mathrm{th,min}$ is achieved if the threshold is reached with the pump laser being exactly resonant with the cavity. This can either approximately be achieved by carefully setting the pump power $P_\mathrm{in}$ or by applying injection locking to the pump laser \citep{Liang2010}. In this ideal case the previous equation simplifies to 
\begin{equation}
\mu_{\mathrm{th,min}}=\sqrt{\frac{\kappa}{D_{\mathrm{2}}}}%
\end{equation}
We note that injection locking may help to operate systems that are at the border line between NMS and MMS dynamics.

\section{Discussion Conclusion}
The surprising and unexpected observation made in many experiments  \citep{Del'Haye2011,Papp2011} that NMS combs generally exhibit narrow, and well defined beat-notes, while MMS combs show multiple and broad beat-notes is explained in this work for the first time and attributed to their different dynamics of formation. While NMS combs can reproduce and transport the initial spacing between pump and first generated side-bands to all subsequently emerging modes via cascaded non-degenerate FWM, this is not the case for MMS combs. Here, the non-commensurability of the spacing $\Delta$ of the primary comb lines with the Kerr-combs native line spacing $\delta$ leads to spectral inconsistencies. These inconsistencies are shown here to give rise to multiple and finally broad RF beat-notes when filling the gaps between the primary lines. 

Our experimental observations in conjunction with the theoretical analysis implies that the relevant figure of merit for the design of low phase noise Kerr-comb generators is the ratio of cavity decay rate (i.e. cavity linewidth) to dispersion $D_{\mathrm{2}}$. Ideally, this ratio should be close to unity, which can either be achieved by reducing the cavity decay rate $\kappa$ or increasing the dispersion $D_{\mathrm{2}}$. Note that this ratio has been achieved in e.g. ref.~\citep{Del'Haye2007}.


Many system such as Si$_3$N$_4$ ring-resonators, posses great freedom in choosing their geometrical design parameters, which allows to strongly influence their dispersion. While influencing the dispersion geometrically is only marginally possible in crystalline resonator systems, like the presented MgF$_{2}$ resonators, this technology platform remains promising when considering comb generation in wavelength regions, where the dispersion is strongly anomalous, i.e. the mid-infrared regime. This is particularly important for the mid-IR regime where Kerr-comb generation has been demonstrated recently in the strong anomalous dispersion regime (above $2500$~nm) with no evidence of multiple or broad beat-notes \citep{Wang2011}.

Another way of influencing the dispersion $D_{\mathrm{2}}$ is increasing the free spectral range of the resonator, as $D_{\mathrm{2}}$ can be interpreted as the dispersion of the resonator integrated over the frequency interval of one FSR. This explains, why systems with high repetition rate are advantageous to generate NMS combs. We note that the condition $\mu\sim1$ is not a strict one. As long as the mismatch between $\mu\cdot\delta$ and $\Delta$ observed in Fig.~\ref{exp1fig}c is smaller then the laser linewidth no multiple beat-notes will be observed. While the analysis presented here is purely frequency domain based, it seems likely that the mechanism described here can also account for recent and unexplained observations made in the time domain \citep{Weiner2011}, where the partial loss of contrast in auto-correlation measurements was observed for combs following the MMS pathway as opposed to those of NMS nature.

\section*{Acknowledgements} 
This work was funded by a Marie Curie IAPP, Eurostars, the Swiss National Science Foundation, the NCCR Nanotera (NTF), and  the DARPA QuASAR program. M.~L.~Gorodetksy acknowledges support of the Dynasty Foundation. The SiN samples were grown and fabricated in the CMI-EPFL nano-fabrication facility. C.W. acknowledges support by a Marie Curie IEF.

\section{METHODS}
\subsection{Multi-heterodyne experiment}
\label{methods:multhet}
 A narrow linewidth, continuous wave $1553$~nm fibre laser is amplified by an EDFA to $80$~mW of power and sent to the MgF$_{2}$ resonator for Kerr-comb generation. The pump laser is offset locked to $20$~MHz below a line of the fibre laser comb. The performance of the lock is verified by a RF frequency counter. The resonator is thermally locked to the pump laser and before each measurement we wait until the system shows negligible thermal drift. After comb generation the generated spectrum is split and into several beams. One beam is sent to an optical spectrum analyser(OSA) for detection of the optical Kerr-comb spectrum, a second one is sent to a fast $45$~GHz photo-diode for detection of the RF beat-note between natively spaced neighbouring comb lines, whenever they are present. The electronic signal generated in the fast photo-diode is down-mixed using the third harmonic of an RF generator at $11.72$~GHz and sent to an electronic spectrum analyser (ESA) and another frequency counter. Note that the pump was attenuated by approx. $30$~dB using a narrow (only affecting the pump laser line) fibre-Bragg grating in transmission. 
Finally, two more beams are combined with the spectrum of a fibre-laser comb possessing a repetition of $250$~MHz. Two tunable optical bandpass filters are used to filter out a narrow spectral region of the combined comb spectra (the $0.8$~nm bandwidth of the filters is much narrower compared to the spacing of the primary lines) at the position of the primary comb lines and sent to two photo-diodes for detecting the multi-heterodyne beat-note between the two combs (which typically exceed a signal-to-noise ratio of $20$~dB in $300$~kHz). The beat-notes frequencies are determined by frequency counters with a gate time of $0.1$~s. The heterodyne beat-notes, in combination with the offset lock of the pump, allow to accurately determine the frequency difference between primary comb lines and pump frequency. Hereto we have assumed that the Kerr-Comb lines do not deviate by more than $250$~MHz from their positions expected based on the FSR of the resonator.

\subsection{Kerr-comb reconstruction}
\label{methods:combreco}
The setup follows the approach developed in ref.~\citep{Del'Haye2009} and develops it further in order to spectrally reconstruct Kerr-comb spectra. To this end a diode laser, scanning over a frequency interval of approx. $4$~THz (from $1550$~nm towards longer wavelength) is split into three beams, two of which serve to frequency calibrate the laser scan. The first calibration beam is combined with a the spectrum of a fibre-laser comb with a repetition of $250$~MHz. The signal is detected by a photo-diode and the resulting RF signal is band-pass filtered around $10$~MHz with a bandwidth of $2$~MHz and continuously recorded by an oscilloscope during the scan. Whenever, the scanning laser sweeps over a position of $10$~MHz below or above a fibre-laser comb line a RF signal passes the bandpass filter, resulting in two markers per comb line in the oscilloscope trace, based on which the instant in time when the laser crossed a a fibre laser comb line can be found. This allows to define a relative detuning of the pump laser as a function of time during the scan. 

To further enhance the precision of the calibration the second beam of the diode laser is coupled to a fibre-loop cavity with a FSR of $\sim10$~MHz and the transmission is recorded simultaneously on another oscilloscope channel. The resonances of the fibre-loop cavity provide additional calibration marker that are used for fine calibration of the laser detuning in between two fibre laser comb lines. Note that effect of fibre dispersion is small (close to zero dispersion) and can be neglected here. Generally, a spectrally local calibration of the FSR of the fibre-loop cavity can be done using two neighbouring fibre-laser comb markers. We estimate the precision of the method to $\leq 10\%$ of the fibre-loop FSR, i.e below $1$~MHz. Note that in principle a much finer calibration grid is possible by decreasing the FSR of the fibre-loop cavity, which however, would require a higher resolution oscilloscope. The main limitation in our measurement comes from the unsteady scanning behaviour of the used diode laser, which we attribute to the mechanics of the stepper-motor driven scan and low frequency noise present in the electric supply network. These observed scan speed variations over intervals as small as the FSR of the fibre loop significantly impact the quality of the spline interpolation between the markers. Important in this regard is that the narrow spacing of the fibre-loop cavity resonances allows to faithfully detect sudden variations in the scan speed of the laser, change of scan direction, as well as mode hops, which would lead to spurious results. 
The third diode laser beam is used to probe the Kerr-comb spectrum by generating and recording RF marker in same fashion as for the fibre-laser comb. As detailed above these markers allow to determine the position of the Kerr-comb lines within the scan. 

Based on the above data of simultaneously recorded calibration markers and Kerr-comb markers a relative frequencies can be assigned to the Kerr-comb lines (cf. Fig.\ref{exp2fig} a,b,c). For reconstruction of the comb lines only $250$~MHz intervals between fibre laser comb lines where the laser scan was unidirectional and smooth are used. To correct for mode hops that happened at some point during the scan we assume, that the position of the reconstructed Kerr-comb lines does not deviate by more then $250$~MHz from an equidistant grid, i.e. we subtract integer multiples of $250$~MHz from the expected equidistant Kerr-comb line position. This assumption is justified, as the residuals after this operation are typically well below $0.1$ in units of the fibre frequency comb's repetition rate of $250$~MHz. Further data quality criteria applied in the reconstruction include the height of the comb calibration markers to allow accurate comb line reconstruction. As the pump laser is offset locked to the fibre-laser, the pump line is most easily identified with the Kerr-comb line reconstructed with the corresponding relative detuning to the next fibre laser comb line. Note that this method can resolve lines not resolvable by the OSA (e.g. close to the pump).

The method described here, in particular employing a low-cost fibre-loop cavity to enhance the precision and robustness of frequency comb assisted diode laser spectroscopy \citep{Del'Haye2009} is of general interest for high-precision spectroscopic experiments. As the fine calibration grid is provided by the fiber-loop cavity, we note that the frequency comb could in principle be replaced by a low-cost gas cell, where precisely known positions of absorption lines can replace the frequency comb lines.

\section*{SUPPLEMENTARY INFORMATION}
\subsection{First Oscillating Mode}

Here we address the question how distant (in terms of mode number difference) the first parametrically generated modes are with respect to the pump mode. Experimentally, two procedures for comb generation exist:

(1) Thermal locking: The pump laser is set to a fixed power and is initially strongly blue detuned, from the pump resonance. This detuning is then slowly reduced, such that more and more light is coupled to the resonance. At some point the parametric threshold is reached and the first side-mode pair is generated depending on the current detuning and power level.

(2) Injection locking: Here, the back reflection of the cavity is used as a feedback to an initially broad gain laser, thereby narrowing the laser linewidth and locking the laser to the cavity, such that the detuning between the pump laser frequency and the hot cavity resonance frequency is equal to zero. With the laser locked exactly onto resonance the power is increased until parametric gain is reached and the first side-mode pair is generated.

We will commence our considerations with finding the solution for the first procedure for comb generation and
then later see that it also contains the solution for the second procedure as a special case.

When a laser power with frequency $\omega_p$ is pumped to a cavity, a system of nonlinear coupled mode equations \cite{Kippenberg2004a,Matsko2005,Chembo2010,Strekalov2010,Chembo2010b,Matsko2010,Matsko2011} can be used to describe the evolution of the mode amplitudes $A_\mu$ normalized such that $|A_\mu|^2$ is the number of quanta in the mode $\mu$. All
mode numbers $\mu$ are defined relative to the pumped mode $\mu=0$). Using the cold cavity eigenmodes with frequencies $\omega_\mu = \omega_0+D_1 \mu+ \frac{1}{2} D_2 \mu^2$ ($D_1$ and $D_2$ correspond to the FSR of the resonator and the difference between
two neighboring FSRs at the center frequency $\omega_0$, respectively) the simplified set of equations read:
\begin{align}
\frac{\partial A_\mu}{\partial t}&=-\frac{\kappa}{2} A_\mu+\delta_{\mu0}\sqrt{\kappa_{\text{ext}}} s e^{-i(\omega_p-\omega_0)t} \nonumber\\
&+ig\!\!\!\!\!\sum_{\mu',\mu'',\mu'''}\!\!\!\!A_{\mu'}A_{\mu''}A^*_{\mu'''}e^{-i(\omega_{\mu'}+\omega_{\mu''}-\omega_{\mu'''}-\omega_\mu)t}.
\end{align}
Here, $\kappa = \kappa_0 + \kappa_{\text{ext}}$ denotes the cavity decay rate as a sum of intrinsic decay rate $\kappa_0$ and coupling rate to the waveguide $\kappa_{\text{ext}}$. We assume without losing generality that the initial phase of the pump is zero while $s = \sqrt{P_{\text{in}}/\hbar\omega_0}$ denotes the amplitude of the pump power $P_{\text{in}}$ coupled to the cavity and $\delta_{\mu 0}$ is Kronecker's delta. The non-linear coupling coefficient
\begin{align}
g&=\frac{\hbar\omega^2_0 cn_2}{n_0^2V_{\text{eff}}}.
\end{align}
describes the cubic non-linearity of the system with the refractive index $n_0$, non-linear refractive index $n_2$, the effective cavity nonlinear volume $V_{\text{eff}}$, the speed of light $c$ and the Planck constant $\hbar$. Physically $g$ denotes a per photon frequency shift of the cavity due to the Kerr nonlinearity. The summation is done for all $\mu',\mu'',\mu'''$ respecting the relation $\mu=\mu'+\mu''-\mu'''$. 

This system may be written in a simplified dimensionless way using scaling proposed in \cite{Matsko2005}: $f=\sqrt{8\eta g/\kappa^2}s$, $d_2=D_2/\kappa$, $\zeta_{\mu}=2(\omega_\mu-\omega_p- \mu D_1)/\kappa=\zeta_0+d_2\mu^2$, $\tau=\kappa t/2$, and use also phase transformation $a_\mu=A_\mu\sqrt{2g/\kappa}e^{-i(\omega_\mu-\omega_p-\mu D_1)t}$ to get rid of time dependences of nonlinear terms. Here $\eta=\kappa_{\text{ext}}/\kappa$ is the coupling strength, which turns to $1/2$ for critical coupling.
\begin{align}
\frac{\partial a_\mu}{\partial \tau}&=-[1+i\zeta_\mu] a_\mu \nonumber\\
&+i\sum_{\mu'\leq \mu''}(2-\delta_{\mu'\mu''}) a_{\mu'} a_{\mu''} a^*_{\mu'+\mu''-\mu}+ \delta_{0\mu} f.
\end{align}
In this form all frequencies, detunings and magnitudes are measured in units of cold cavity resonance linewidth so that $|a_\mu|^2=1$ corresponds to the nonlinear mode pulling on one cold cavity resonance (thresholds for both single mode bistability and degenerate oscillations \cite{Chembo2010}). This system we used for theoretical analysis of the comb formation and for numerical simulations \cite{ArchiveGHK}. To approximate the modes' field distribution and eigenfrequencies the model of spheroidal cavity was used \cite{Gorodetsky2009,Gorodetskybook}.

In order to determine the mode numbers $\pm \mu_{\text{th}}$, of the first modes generated at parametric threshold it is sufficient to consider a three mode system of $a_0$, $a_{+\mu}$, and $a^*_{-\mu}$ \cite{Chembo2010}, i.e. pump, signal, and idler mode. The steady  state for the pump mode when no side-modes are excited (without nonlinear terms) gives:
\begin{align}
(\zeta_0-|a_0|^2)^2|a_0|^2+|a_0|^2=f^2. \label{pump}
\end{align}
Characteristic equation for the linearized system for $a_{+\mu}$, and $a^*_{-\mu}$ near zero when only pump mode is excited:
\begin{align}
\frac{\partial a_{+\mu}}{\partial \tau}&=-[1+i\zeta_\mu+2i|a_0|^2] a_{+\mu}+ia_0^2 a^*_{-\mu},\nonumber\\
\frac{\partial a^*_{-\mu}}{\partial \tau}&=-[1-i\zeta_\mu-2i|a_0|^2] a^*_{-\mu}-ia^{*2}_0 a_{+\mu}.
\end{align}
provides
\begin{align}
\lambda=-1\pm \sqrt{|a_0|^4-(\zeta_\mu-2|a_0|^2)^2}. \label{lambda}
\end{align}
The value $G= \Re(\lambda+1) \kappa$ denotes the gain of the side-modes \cite{Kippenberg2004a}, excited if $G>\kappa$. 
\begin{align}
G=\sqrt{\kappa^2(P_{\text{abs}}/P_{\text{th}})^2-4(\omega_0-\omega_p+\mu^2 D_2-\kappa P_{\text{abs}}/P_{\text{th}})^2}, \label{lambda}
\end{align}
where 
\begin{align}
P_{\text{th}}=\frac{\kappa^2n_0^2V_{\text{eff}}}{8\eta\omega_0cn_2}
\end{align}
is threshold power for bistability and $P_{\text{abs}}$ is the power absorbed by the cavity from the waveguide. Combining this condition and  (\ref{pump}) we obtain the equation determining the detuning at threshold:
\begin{align}
&\zeta_0=|a_0|^2-\sqrt{f^2/|a_0|^2-1},
\end{align}
which when inserted into (\ref{pump}) yields for threshold gain: 
\begin{align}
&\sqrt{f^2/|a_0|^2-1}-d_2 \mu_{\text{th}}^2+|a_0|^2-\sqrt{|a_0|^4-1}=0
\end{align}
or using the smallest possible $|a_0|^2=1$ when the last radical is real:
\begin{align}
\mu_{\text{th}}= \sqrt{\frac{1}{d_2}(\sqrt{f^2-1}+1)}.
\end{align}
Or transforming dimensionless units back to physical units: 
\begin{align}
\boxed{\mu_{\text{th}}\simeq\sqrt{\frac{\kappa}{D_2}\left(\sqrt{\frac{P_{\text{in}}}{P_{\text{th}}}-1}+1\right)}}
\nonumber\\
\end{align}
The threshold power $P_{\text{th}}$ which corresponds to $|a_{0,\text{th}}|=1$, the normalized energy in the pumped mode when first sidebadns can appear and also a threshold for bistability (see also \cite{Chembo2010}), however does not correspond to a minimum input power when hyperparametric generation may start. This minimum power threshold, numerically calculated in \cite{Matsko2005} may be found explicitly and corresponds to $|a_{0,\text{min}}|=2/\sqrt{3}$.

The minimum $\mu_{\text{th}}$ obtained at threshold $f=1$ at the edge of hysteretic nonlinear curve may be found as: 
\begin{align}
\boxed{\mu_{\text{th,min}}=
\sqrt{\frac{\kappa}{D_2}}},
\end{align}
which only depends on the ratio of cavity decay rate to second order cavity dispersion $D_2$. 

In analogous way a condition for generating intermediate side-modes (cf. Fig. 2f), with nondegenerate FWM process $0+\mu=\mu'+\mu''$ may be written, when a strong initial side-mode is already excited with linearized set of equations (and analogous for complex conjugate):
\begin{align}
\frac{\partial a_{\mu'}}{\partial \tau}&=-[1+i\xi_{\mu'}+2i|a_0|^2+2i|a_\mu|^2] a_{\mu'}+2ia_0a_\mu a^*_{\mu''},\nonumber\\
\frac{\partial a^*_{\mu''}}{\partial \tau}&=-[1-i\xi_{\mu''}-2i|a_0|^2-2i|a_\mu|^2] a^*_{\mu''}-2ia_0^*a^*_\mu a_{\mu'}. \label{intermediate}
\end{align}
In fact this is a simplified set of equations, as the whole set should include mixing not between 4 but 7 comb lines (0, $\pm\mu_1$, $\pm\mu_2$, $\pm\mu$) which can be described by 4 nonlinearly coupled linearized equations.

The eigenvalues for (\ref{intermediate}) are:
\begin{align}
\lambda=-1+\frac{i}{2}(\xi_{\mu''}-\xi_{\mu'}) \nonumber\\
\pm 2\sqrt{|a_0|^2|a_\mu|^2-(|a_0|^2+|a_\mu|^2-(\xi_{\mu'}+\xi_{\mu''})/4)^2}
\end{align}
As $\xi_{\mu''}+\xi_{\mu'}=2\zeta_0+d_2(\mu^{'2}+(\mu-\mu')^{'2})$ the gain has maximum when $\mu'=\mu/2$ and minum for $\mu'=1$, which means that secondary combs depending on the values of primary amplitudes detuning can appear in the centers of the gaps between initial multi-FSR side-modes start or adjacent to primary lines. It is essential that if $\mu'\neq\mu''$ these secondary side-modes provide two sidebands detuned from equidistant comb-lines to $(\xi_{\mu''}-\xi_{\mu'})\kappa/4 = \pm D_2(\mu^{''2}-\mu^{'2})/4$, giving rise to RF beat note peaks at doubled frequency.

\subsection{Optical Comb Spectra and RF Beat-Notes}
\begin{figure*}[ptbh]
\begin{center}
\includegraphics[width=0.8\textwidth]{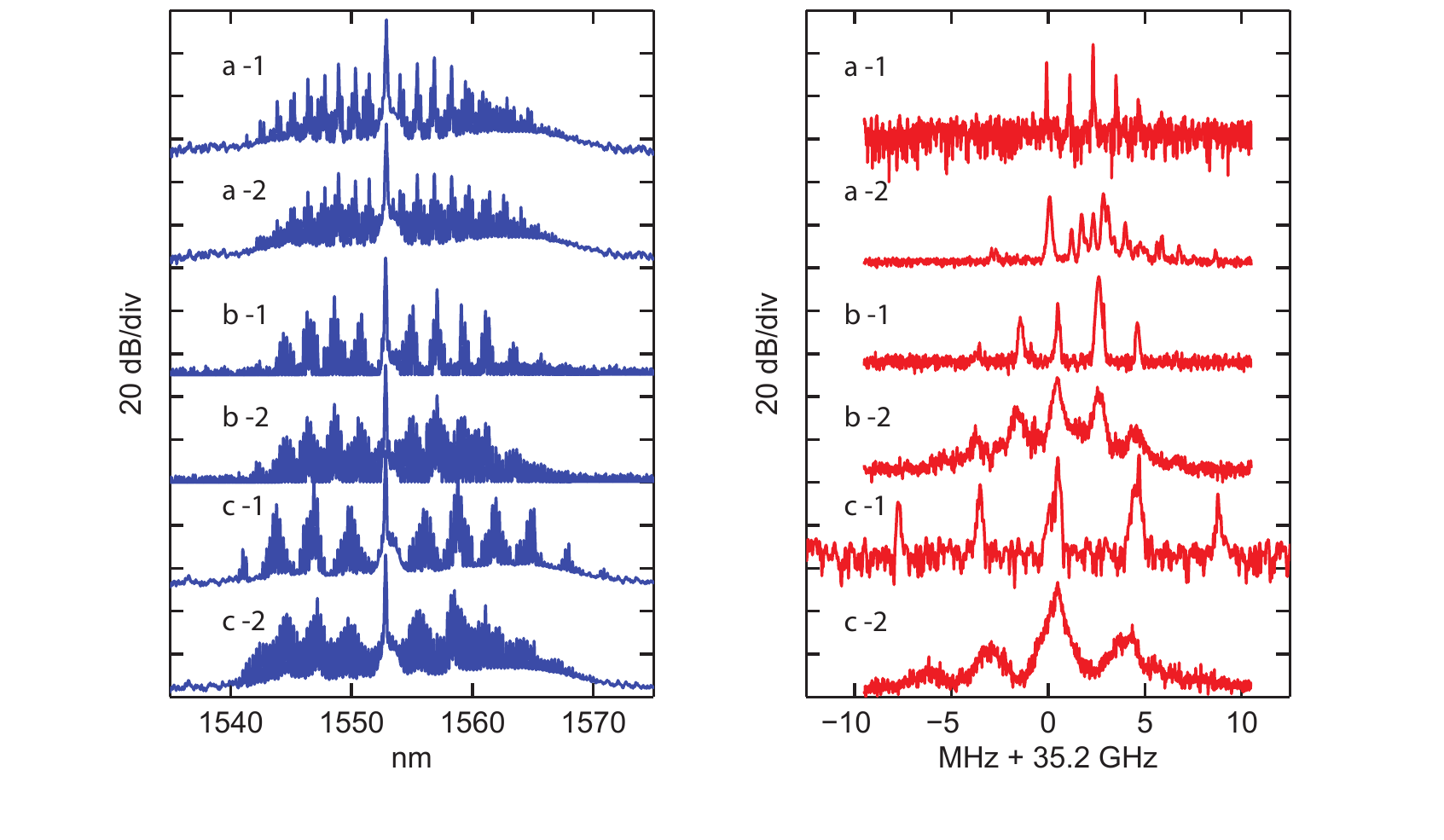}
\end{center}
\caption{Optical Kerr-comb spectra (blue) and corresponding radio-frequency (RF) beat-notes (red) generated in a $35$~GHz MgF$_{2}$ resonator by pumping different optical mode families. Each comb spectrum (a,b,c) is shown in an early state of its evolution (1) and after increasing the circulating power by reducing the the detuning of the pump laser with respect to the pump resonance (2).}
\label{combAndRep}%
\end{figure*}

\begin{figure*}[htbp]
\begin{center}
\includegraphics[width=0.4\textwidth]{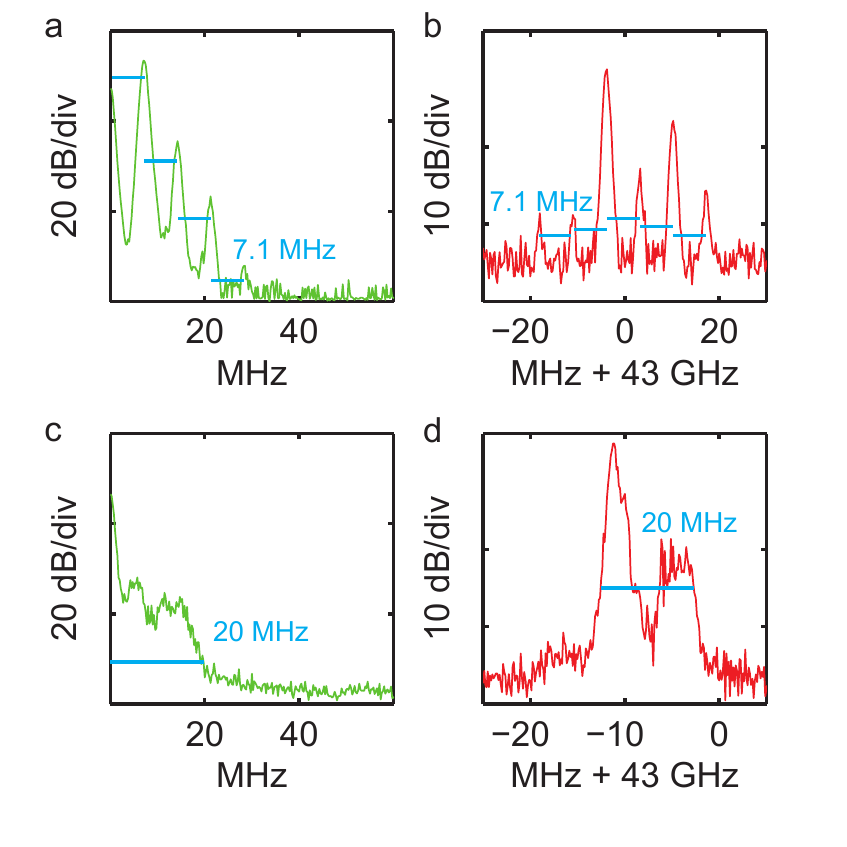}
\end{center}
\caption{\textbf{a. } Low frequency RF spectrum of the transmitted pump power during comb formation and \textbf{b.} high frequency beat-note between neighbouring comb lines recorded in a resolution bandwidth of $300$~kHz. \textbf{c,d.} Same as (a,b) for a different comb state. The measurements demonstrates that the RF beatnotes find their correspondence in the low frequency fluctuations of the transmitted pump power}
\label{repAndTransLowRF}%
\end{figure*}

To illustrate the complexity of the underlying physics different comb spectra and their corresponding RF beat-notes are shown in Fig.~\ref{combAndRep}. While the observed beat-notes are multiple in all cases they can be narrow or broad, embedded in a unresolved noise pedestal, and highly-asymmetric in amplitude and spacing. In Fig.~\ref{combAndRep}a and c illustrate the impact of resonator dispersion as two different mode families were used for comb generation. Fig.~ref{combAndRep}b,c shows the effect of pump power (b: low, c: high) where the primary side-bands are moving away from the pump with higher pump power.


\subsection{High Frequency and Low Frequency Radio Frequency Signal}
The difference between the high frequency ($\gtrsim 10$~GHz) multiple RF beat-notes can be observed in the electronic low frequency regime. This can be in the form of narrow single or multiple peaks or broad band noise like features. In the data obtained in a crystalline $35$~GHz MgF$_{2}$ resonator the difference frequencies between multiple RF beat-notes are in the range form one to several tens of Megahertz. In the Si$_3$N$_4$ system (cf. main manuscript), however, these difference frequencies have been observed as high as several hundreds of Megahertz.


\end{document}